%% file: main.tex
\documentclass[sigconf,screen,balance=false]{acmart}

\setcopyright{none}
\renewcommand\footnotetextcopyrightpermission[1]{}
\usepackage{booktabs}
\usepackage{multirow}
\usepackage{tabularx}
\usepackage{enumitem}
\usepackage{xcolor}
\usepackage{graphicx}
\definecolor{findingborder}{HTML}{46545C}
\definecolor{findingfill}{HTML}{F2F5F7}
\definecolor{exampleframe}{HTML}{B8C7D1}
\definecolor{examplefill}{HTML}{F7F9FA}
\newcounter{finding}
\newcommand{\finding}[1]{%
  \refstepcounter{finding}%
  \par\vskip 5pt\noindent
  \begingroup
  \setlength{\fboxsep}{7pt}%
  \fcolorbox{findingborder}{findingfill}{%
    \parbox{\dimexpr\columnwidth-2\fboxsep-2\fboxrule\relax}{%
      \small\textbf{Finding \thefinding.}~#1}}%
  \par\vskip 5pt
  \endgroup}
\newcounter{exampleblock}

\title{Skill or Threat? Risks and Scanner Gaps in the Large Language Model Skill Ecosystem}

\author{Xinghao Gao}
\affiliation{\institution{East China Normal University}\city{Shanghai}\country{China}}
\email{zewang0217@outlook.com}

\author{Xinhe Zhang}
\affiliation{\institution{East China Normal University}\city{Shanghai}\country{China}}
\email{zxh0826@njfu.edu.cn}

\author{Yanxiang Mao}
\affiliation{\institution{East China Normal University}\city{Shanghai}\country{China}}
\email{kutbass@163.com}

\author{Jiongchi Yu}
\affiliation{\institution{Nanyang Technological University}\city{Singapore}\country{Singapore}}
\email{jiongchiyu@acm.org}

\author{Xiaoyu Zhang}
\affiliation{\institution{Nanyang Technological University}\city{Singapore}\country{Singapore}}
\email{joshingrain@gmail.com}

\author{Kaixuan Li}
\affiliation{%
  \institution{Nanyang Technological University}
  \city{Singapore}
  \country{Singapore}
}
\email{kaixuan.li@ntu.edu.sg}

\author{Ming Hu}
\authornote{The corresponding author.}
\affiliation{\institution{ East China Normal University}\city{Shanghai}\country{China}}
\email{mhu@sei.ecnu.edu.cn}

\author{Geguang Pu}
\affiliation{\institution{East China Normal University}\city{Shanghai}\country{China}}
\email{ggpu@sei.ecnu.edu.cn}

\begin{abstract}
Agent Skill scanners use different risk categories, making it difficult to
tell whether reports for the same Skill describe the same behavior or provide
additional information. We synthesize threat classifications from 34 academic
papers and 15 industry reports, standards, and frameworks, along with
benchmark labels, scanner categories, and rules, into a taxonomy organized by
Source, Mechanism, and Target. We compare public reports from four scanners
for 386 marketplace Skills. We also scan 350 Skills labeled as malicious by
MalSkillBench with three open-source scanners and compare reports for the
317 Skills with results from all three.

Native categories often omit threat information or overlap in scope.
Counting each (Skill, Mechanism) pair once, 79.9\% of marketplace pairs and
82.4\% of malicious-Skill pairs appear in only one scanner's reports. Among
the analyzed malicious Skills, 85.5\% have at least two Mechanisms and
53.3\% have at least two Targets. The taxonomy and evidence from the original
reports help distinguish different descriptions of the same operation from
additional reported risks and provide a common basis for combining reports.

\end{abstract}


\begin{document}
\maketitle
\pagestyle{plain}
\raggedbottom

\input{sections/01-introduction}
\input{sections/03-related-work}
\input{sections/04-study-overview}
\input{sections/08-results-rq1}
\input{sections/09-results-rq2}
\input{sections/10-results-rq3}
\input{sections/11-discussion}
\input{sections/12-threats-ethics}
\input{sections/13-conclusion}

\bibliographystyle{ACM-Reference-Format}
\bibliography{references}
\appendix
\input{sections/15-taxonomy-relations-appendix}
\input{sections/16-platform-collection-appendix}
\input{sections/17-full-statistics-appendix}
\input{sections/14-appendix-corpora}

\input{sections/18-taxonomy-sources-appendix}
\end{document}

%% file: sections/01-introduction.tex
\section{Introduction}
\label{sec:introduction}

Agent Skills organize task instructions and scripts into reusable workflows that extend the capabilities of AI agents~\cite{agentskillsspec}.
These materials influence an agent's tool use, file access, and external communication, creating entry points for malicious instructions and sensitive information leakage~\cite{skillinject2026,greshake2023indirect}.
To identify these risks, vendors and Skill platforms have introduced security scanners.

Scanners differ in their risk focus and detection approaches, so combining their outputs can provide broader evidence for security review~\cite{clawhub2026}.
Skill platforms such as skills.sh and ClawHub already use multiple scanners to assess the same Skill~\cite{skillsshaudits,clawhubmultiscanner}.
The skills.sh platform displays security reports from Agent Trust Hub, Socket, and Snyk~\cite{skillsshaudits}.
ClawHub combines outputs from static analysis, VirusTotal, and SkillSpector to inform publication review~\cite{clawhubmultiscanner}.
Both platforms serve large Skill catalogs.
The skills.sh directory listed over 1.3 million Skills on 18 October 2026~\cite{skillsshcatalog}.
ClawHub's registry snapshot from May 2026 already contained over 67,000 latest public Skill versions~\cite{clawhub2026}.

However, multiple reports do not automatically yield a coherent security assessment.
Scanners often produce different risk descriptions and security decisions for the same Skill.
Without establishing how their findings relate, users and platforms cannot tell which reported risks overlap and which additional risks a combination of scanners reveals.
Understanding these differences is necessary for comparing scanners, combining scanners, and identifying issues that require security review.

Prior work examines scanner disagreement from several perspectives.
ClawHub Security Signals compares the Skills flagged by multiple scanners and examines attack surfaces to explain disagreement~\cite{clawhub2026}.

SkillsMetric studies static detection across attack types~\cite{skillsmetric2026}.
For scanner evaluation, MalSkillBench, MaliciousSkillBench, and SkillTrustBench provide attack samples and risk labels~\cite{malskillbench2026,maliciousskillbench2026,skilltrustbench2026}.
Together, these efforts provide a basis for studying scanner differences.

\begin{figure*}[t]
  \centering
  \includegraphics[width=\textwidth]{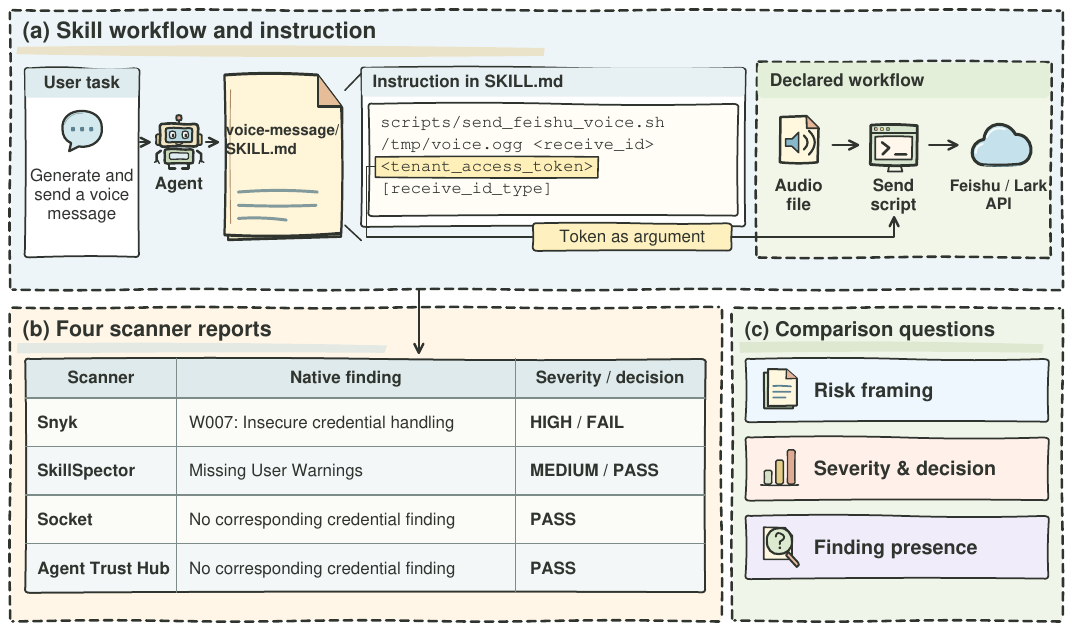}
  \caption{Scanner reports for the same \emph{voice-message} Skill.
  (a) The Skill workflow and preserved command pass an access token as a script
  argument. (b) Four public reports retain their native finding names,
  severity assessments, and decisions. (c) Comparing the reports requires
  explaining their risk framing, severity and decisions, and whether a
  corresponding finding is present.}
  \Description{The voice-message Skill workflow and command pattern are shown alongside four public scanner reports. Snyk reports insecure credential handling with HIGH severity and FAIL. SkillSpector reports Missing User Warnings with MEDIUM severity and PASS. Socket and Agent Trust Hub report PASS without a corresponding credential finding.}
  \label{fig:voice-message-case}
\end{figure*}

These studies examine flagged Skills, detection boundaries, and benchmark labels~\cite{clawhub2026,skillsmetric2026,malskillbench2026,maliciousskillbench2026,skilltrustbench2026}.
Their comparisons leave the semantic relationships between different scanners' risk categories unresolved.
Consequently, they leave open whether divergent reports describe the same risks, partially overlapping risks, or different risks.
The same reports can contain both shared findings and findings unique to a scanner.
We therefore need to distinguish differences in category definitions from differences in reported risk content.
These distinctions help explain scanner disagreement and the information provided by combining reports.

A central obstacle is the lack of a common risk classification.
Scanners, academic studies, and industry frameworks use their own categories for Skill risks.
These categories differ in name and may describe different aspects of a threat.
Some describe how an attack operates, while others describe its security consequences.
Still others group several behaviors into a broad risk category.
Comparing category names directly can count different descriptions of the same risk as additional findings and overstate the complementary information provided by combining scanners.
Conversely, merging similarly named categories can conceal differences in their scope.
We must therefore establish relationships between categories before comparing the shared and additional information in individual reports.

The \emph{voice-message} Skill in Figure~\ref{fig:voice-message-case} illustrates this problem.
We collected four public security reports for this Skill as part of our marketplace collection.
The SkillSpector report comes from ClawHub, while the Snyk, Socket, and Agent Trust Hub reports come from skills.sh.
Section~\ref{sec:corpora} describes the collection procedure.
The Skill instructs an agent to generate audio and invoke the Feishu/Lark application programming interface (API) through a script.
The command passes an access token as an argument.
Snyk Agent Scan reports a credential-handling risk with high severity and assigns a blocking decision~\cite{snykissuecodes}.
SkillSpector reports missing user warnings and passes the Skill with a medium-severity assessment~\cite{skillspectordocs}.
Socket and Agent Trust Hub pass the Skill without a corresponding credential finding in their reports~\cite{socketalerttypes,agenttrusthubdocs}.

These reports concern related operations but give different risk descriptions and decisions.
Counting which scanners raise an alert does not explain what the reports share.
Treating different finding names as separate issues also fails to identify their common content.

We address this gap by developing a common risk representation across categories and applying it to scanner reports.
We first examine the threat information expressed by native category definitions and the semantic relations between them.
We then compare the Mechanisms reported by different scanners for the same Skill.
Finally, we combine reports with Skill content to characterize the Sources, Mechanisms, and Targets of malicious Skills.

We review threat classifications from 34 academic papers and 15 industry reports, standards, and frameworks to construct the taxonomy.
Representative sources include academic threat analyses and industry risk frameworks~\cite{maloyan2026coding,li2026secureskills,mitreatlas,owaspllm}.
Appendix~\ref{app:taxonomy-sources} lists the academic and industry sources and the classification perspectives examined.
We integrate these classifications with native category and rule definitions and existing benchmark labels to develop a unified Skill risk taxonomy.
The detailed inventory used to examine category information covers four scanners, one security standard, and one threat rule system.
The four scanners are SkillSpector, Snyk Agent Scan, Socket, and Cisco Skill Scanner~\cite{skillspectordocs,snykissuecodes,socketalerttypes,ciscotaxonomy}.
The security standard is Gen Digital's \emph{An Open Standard for AI Agent Runtime Safety} (AARTS), and the threat rule system is Agent Threat Rules~\cite{aartsstandard,atrdocs}.
Additional benchmark and scanner category definitions support the category-relation and report analyses.

The taxonomy has three facets, Source, Mechanism, and Target.
They capture where malicious influence enters, how it operates, and its security consequences.
We use this representation to establish semantic mappings between categories while preserving their shared content and differences in scope.
These mappings provide a common basis for comparisons across benchmarks and scanners.

We investigate three research questions.

\begin{itemize}[leftmargin=*]
  \item \textbf{RQ1.} What threat information do native risk categories express, and what semantic relations exist between them?
  \item \textbf{RQ2.} Which Mechanisms are shared or different in scanner reports for the same Skill?
  \item \textbf{RQ3.} Which Sources, Mechanisms, and Targets characterize malicious Skills?
\end{itemize}

We apply the taxonomy to native category relations and benchmark risk labels~\cite{malskillbench2026,maliciousskillbench2026,skilltrustbench2026}.
We compare public reports from four scanners for 386 marketplace Skills.
We also run three open-source scanners on 350 Skills labeled as malicious by MalSkillBench, of which 317 have traceable reports from all three scanners and enter the comparative analysis~\cite{skillspectordocs,snykissuecodes,socketalerttypes,agenttrusthubdocs,ciscotaxonomy,caterpillardocs,malskillbench2026}.
Together, the two analyses cover six distinct scanners, with SkillSpector present in both.
Section~\ref{sec:overview} explains the materials and scanner selection for each analysis.

Our analysis shows that native categories can leave threat information
unspecified and have overlapping or nested semantic scopes.
The taxonomy makes these relationships and information gaps explicit.
Scanner reports frequently describe different Mechanisms for the same Skill,
while alignment also reveals shared Mechanisms and additional findings.
Counting each (Skill, Mechanism) pair once, 79.9\% of marketplace pairs
and 82.4\% of malicious-Skill pairs appear in only one scanner's reports.
Among the malicious Skills, 85.5\% have at least two Mechanisms and 53.3\%
have at least two Targets. A single native category can also encompass
different Mechanisms. These results show how common risk descriptions and
instance evidence help explain report differences and complex threats.

Our main contributions are as follows.

\begin{itemize}[leftmargin=*]
  \item \textbf{A unified Skill risk taxonomy.} We synthesize existing threat classifications, scanner categories, and benchmark risk labels to establish a common Source, Mechanism, and Target representation with semantic mappings across sources.
  \item \textbf{An analysis of native risk categories.} We examine the threat information expressed by category definitions and their semantic relations, identifying shared content, overlapping scopes, and information gaps in scanner and benchmark classifications.
  \item \textbf{An empirical study of scanner reports and malicious Skills.} We compare shared and additional Mechanisms in public marketplace reports and local scan results, and characterize the Sources, Mechanisms, and Targets of malicious Skills.
\end{itemize}

These findings help interpret and integrate scanner reports.
They also motivate a shared risk taxonomy and reporting conventions.
We call on scanner vendors and Skill platforms to adopt consistent risk descriptions and retain supporting behavioral evidence.
This would reduce ambiguity across reports and support scanner improvement, scanner combination, and security review.

%% file: sections/03-related-work.tex
\section{Related Work}
\label{sec:related-work}

Research on indirect prompt injection shows how untrusted content and tool
responses can redirect model behavior~\cite{greshake2023indirect,debenedetti2024agentdojo}.
Studies of Agent Skills examine vulnerabilities, malicious instructions and
code, and scanner evasion~\cite{skillscan2026,skillprobe2026,skillinject2026,locatejudge2026,skillsmetric2026,cloakdetonate2026,colluskill2026}.
MalSkillBench, MaliciousSkillBench, and SkillTrustBench provide attack samples
and risk labels for scanner evaluation~\cite{malskillbench2026,maliciousskillbench2026,skilltrustbench2026}.
We use threat descriptions and labels from these studies to examine how
their classifications relate to scanner categories.

Security taxonomies organize threats by different criteria.
Attack taxonomies distinguish vectors, targets, and vulnerabilities, while
supply-chain taxonomies organize attacks across development and distribution
stages~\cite{hansman2005taxonomy,ladisa2023supplychain}.
Agent-focused classifications examine delivery vectors, operating mechanisms,
architectural layers, and security consequences~\cite{maloyan2026coding,li2026secureskills,chu2026lasm,owaspagentic}.
Work on cyber threat intelligence also examines taxonomies and standards for
sharing threat descriptions~\cite{mavroeidis2017cti}.
These approaches inform our representation of complementary threat facets
and the relationships between native categories.

Studies of traditional static analyzers compare detection, agreement, and
precision~\cite{lenarduzzi2023critical}.
Beller et al. map tool-specific warnings to a common General Defect
Classification (GDC) to compare tools and their
configurations~\cite{beller2016state}.
Lipp et al. and Charoenwet et al. map warnings to Common Weakness Enumeration
(CWE) categories and examine tool combinations in vulnerability detection
and secure code review~\cite{lipp2022staticanalyzers,charoenwet2024securecodereview}.
These studies demonstrate the role of common categories in tool comparison.
We examine semantic relationships between native Skill risk categories and
use them to compare the Mechanisms expressed in reports for the same Skill.

Within the Skill ecosystem, ClawHub Security Signals compares the overlap in
Skills flagged by VirusTotal, static analysis, and SkillSpector and examines
attack surfaces to explain their disagreement~\cite{clawhub2026,virustotaldocs,skillspectordocs}.
It also groups observed evidence using risk themes from the Open Worldwide
Application Security Project (OWASP) and analyzes scanner signals alongside
platform decisions~\cite{clawhub2026,owaspfoundation,owaspagentic}.
We extend the comparison to semantic relationships between native categories
and to shared and additional Mechanisms within individual Skill reports.
We also combine report evidence with Skill content to characterize the
Sources, Mechanisms, and Targets of malicious Skills.

%% file: sections/04-study-overview.tex
\section{Study Design and Methodology}
\label{sec:overview}
\suppressfloats[t]

\begin{figure*}[t]
  \centering
  \includegraphics[width=\textwidth]{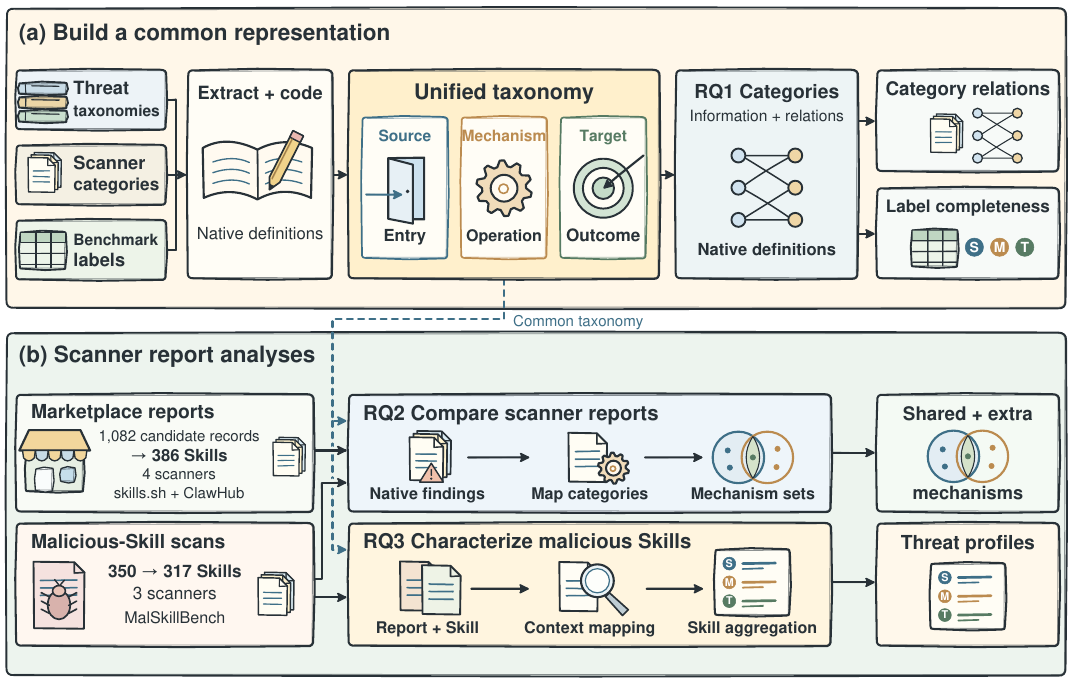}
  \caption{Study overview. (a) Threat taxonomies, scanner categories, and
  benchmark labels inform the Source-Mechanism-Target taxonomy and category
  analysis (RQ1). (b) Native category mappings support scanner-report
  comparison (RQ2), while report and Skill context support malicious-Skill
  characterization (RQ3). Solid arrows indicate processing and dashed arrows
  indicate use of the common taxonomy.}
  \Description{Two horizontal paths summarize the study. The upper path
  connects category sources, extraction and coding, the common taxonomy, and
  category analysis. The lower path connects marketplace and malicious-Skill
  reports to scanner-report comparison and context-informed threat profiles.}
  \label{fig:study-overview}
\end{figure*}

We organize native risk categories into a Source, Mechanism, and Target taxonomy
and apply this representation to scanner reports for marketplace Skills and
Skills labeled malicious by MalSkillBench. This section describes the study
materials, taxonomy construction, report preparation, and analysis methods.
Figure~\ref{fig:study-overview} summarizes taxonomy construction and the three
analyses built on it.

\subsection{Scanners and Category Sources}

The study materials include native category and rule definitions, benchmark
labels, and scanner reports for individual Skills. Definitions and labels
describe categories, whereas a reported finding describes a particular
instance. We relate both to the common taxonomy.
Table~\ref{tab:evidence-units} summarizes these materials and their uses.

\begin{table}[tbp]
  \caption{Study materials, sources, and analytical uses.}
  \label{tab:evidence-units}
  \centering
  \small
  \renewcommand{\arraystretch}{1.05}
  \begin{tabularx}{\columnwidth}{@{}>{\raggedright\arraybackslash}p{0.56\columnwidth}@{\hspace{7pt}}>{\raggedright\arraybackslash}X@{}}
    \toprule
    Material and sources & Analytical use \\
    \midrule
    \textbf{Native category and rule definitions}\newline Category documentation for Cisco, Snyk, Socket, and SkillSpector, together with AARTS and Agent Threat Rules standards or catalogs & Construct the taxonomy and category mappings. Examine definitions and applicability. \\
    \textbf{Benchmark risk labels}\newline MalSkillBench, MaliciousSkillBench, and SkillTrustBench & Map native labels to the taxonomy. Analyze category relations and threat representation. \\
    \textbf{Public marketplace reports}\newline SkillSpector, Snyk, Socket, and Agent Trust Hub reports collected from skills.sh and ClawHub & Map reported findings and compare reports for the same Skill. \\
    \textbf{Reports for malicious Skills}\newline Cisco, SkillSpector, and Caterpillar scans of 350 Skills selected from MalSkillBench & Compare scanners and characterize the threats described for malicious Skills. \\
    \bottomrule
  \end{tabularx}
\end{table}

We extract native definitions from official scanner documentation, security
standards, and rule catalogs~\cite{ciscotaxonomy,snykissuecodes,socketalerttypes,skillspectordocs,aartsstandard,atrdocs}.
We combine these with academic and industry threat classifications and
benchmark labels to construct the taxonomy~\cite{maloyan2026coding,li2026secureskills,mitreatlas,owaspllm,malskillbench2026,maliciousskillbench2026,skilltrustbench2026}.
Appendix~\ref{app:taxonomy-sources} provides the complete academic and
industry source inventory and identifies the classification perspectives
examined in each source.
The named category sources form the detailed inventory used to examine
threat information in native definitions. Category-relation and report
analyses also use additional benchmark and scanner definitions.
We collect public audit pages and their linked scanner reports from skills.sh
and ClawHub using scripts~\cite{skillsshaudits,clawhubdocs}.
At collection time, the platforms exposed six outputs, namely SkillSpector,
Snyk, Socket, Agent Trust Hub, Static Analysis, and VirusTotal.
We compare the first four.
The platform output named Static Analysis showed limited variation in the
collected samples. VirusTotal primarily provides general maliciousness and
reputation signals for files, uniform resource locators (URLs), and domains~\cite{virustotaldocs}, and its
platform output also showed limited variation in these samples.

For the 350 malicious Skills selected from MalSkillBench, we run Cisco Skill
Scanner, NVIDIA SkillSpector, and Caterpillar~\cite{ciscotaxonomy,skillspectordocs,caterpillardocs}.
All three are open source and support Skill file analysis.
Community visibility was a further consideration in selecting Cisco and
SkillSpector. Caterpillar provides a rule-based reference through its
\texttt{offline} mode, which uses built-in pattern matching without calling a
large language model~\cite{caterpillardocs}.

\subsection{Taxonomy Construction}
\label{sec:taxonomy-method}

Existing categories organize risks around entry points, execution operations,
security consequences, weaknesses, or carriers. For example, supply chain
describes how a risk is introduced, prompt injection describes how model
behavior is influenced, and data exfiltration describes an attack
objective~\cite{ladisa2023supplychain,greshake2023indirect,li2026secureskills}.
We extract category names, definitions, and hierarchical relationships,
preserving their sources and evidence locations. We compare related
descriptions and group those that answer the same security question.
Two authors independently coded the material for taxonomy construction,
and a third party adjudicated disagreements.

Multidimensional attack descriptions have precedents in security research.
Hansman and Hunt distinguish attack vector, target, vulnerability, and
payload~\cite{hansman2005taxonomy}.
In a study of coding assistants spanning Skills, tools, and protocols,
Maloyan and Namiot separately organize delivery vectors, attack modalities,
and propagation behavior, and discuss attack objectives in their threat
model~\cite{maloyan2026coding}.
Drawing on this approach, we focus on where a threat enters, how it operates,
and what security consequence it pursues or causes.

The classification purpose and facet-selection criteria draw on
taxonomy-development guidance~\cite{nickerson2013taxonomy}.
We select core facets according to whether their questions recur across
sources, express distinguishable information, and support report comparison
and Skill description using native evidence. Source records the trust origin
or input boundary through which malicious influence enters. Mechanism records
how an attack operates, including manipulation of instructions, code
execution, dependencies, permissions, or state. Target records the security
consequence explicitly pursued or caused. We denote these facets by $S$, $M$,
and $T$ in the analysis definitions.

These aspects can vary separately. Information theft may be pursued through
instructions that induce a model to disclose data or through code that reads
and transmits it. Malicious instructions may arrive with a distributed Skill
or in external documents read during execution. Recording source, mechanism,
and target separately preserves these distinctions. Carrier, weakness,
lifecycle, severity, and product disposition remain auxiliary attributes.

\begin{figure*}[t]
  \centering
  \includegraphics[width=\textwidth]{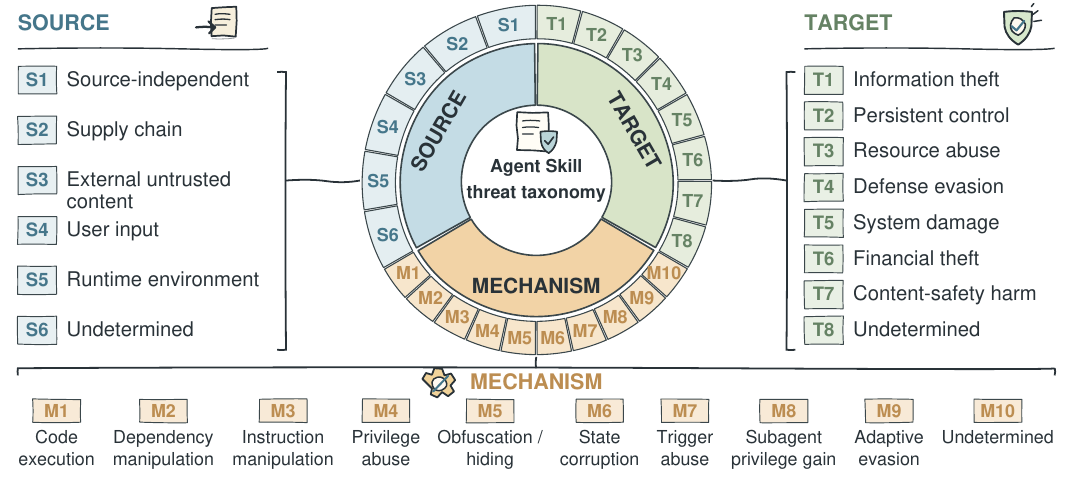}
  \caption{\textbf{Source-Mechanism-Target threat taxonomy.}
  The circular diagram and surrounding lists show all facet values with
  matching display codes. S6, M10, and T8 denote \emph{Undetermined}.
  Multiple supported values may be retained within a facet. Colors and unarrowed
  connectors identify facet groups. Sector area does not encode category
  counts or threat prevalence.}
  \Description{A circular taxonomy with Source values listed on the left,
  Target values on the right, and Mechanism values in one horizontal row below.
  Matching codes link every listed value to its ring segment.}
  \label{fig:shared-threat-space}
\end{figure*}

Within each facet, we organize values from native definitions, merge
semantically consistent descriptions, and use trial mappings to examine
compound categories, ambiguous meanings, and descriptions that existing
values cannot express. These cases guide refinement of the values and their
boundaries. A facet may retain multiple evidence-supported values.
A facet not expressed by a definition remains \emph{undetermined}.
Definitions that explicitly impose no source restriction are recorded as
\emph{source-independent}, and non-threat controls are recorded separately.
Information supplied by a particular report and Skill is retained in an
instance record without changing the native category mapping.
Appendix~\ref{app:taxonomy-relations} describes the definitions, decision
boundaries, examples, and detailed mapping records.
Figure~\ref{fig:shared-threat-space} lists the facet values and their display
codes.

\subsection{Datasets and Report Preparation}
\label{sec:corpora}

Marketplace collection targets Skills listed on both skills.sh and ClawHub.
Our scripts search the two markets and establish cross-platform links using
names, authors, repository information, and preserved \texttt{SKILL.md}
content. This produces 1,082 candidate records, for which we collect public
audit pages and linked reports. The platforms do not provide every scanner's
report for every Skill, so some candidates lack required reports.
We retain Skills with traceable reports from all four study scanners,
yielding 386 Skills for multi-scanner comparison.
Appendix~\ref{app:platform-collection} describes candidate discovery, matching,
and report inclusion.

We select malicious Skills using MalSkillBench's upstream behavior labels.
The selection retains samples from less frequent labels and draws the
remaining samples from the most frequent label. We retain the upstream
malicious labels and supply the preserved \texttt{SKILL.md} files to the three
open-source scanners. Of the 350 selected samples, 317 have traceable reports
from all three scanners. Table~\ref{tab:comparison-cohorts} summarizes the two
collections.

\begin{table}[t]
  \caption{Collection sizes and fixed multi-scanner comparison cohorts.}
  \label{tab:comparison-cohorts}
  \centering
  \small
  \setlength{\tabcolsep}{3pt}
  \begin{tabularx}{\columnwidth}{Xrr}
    \toprule
    Collection & Collected & Compared Skills \\
    \midrule
    Marketplace candidates & 1,082 records & 386 \\
    MalSkillBench & 350 samples & 317 \\
    \bottomrule
  \end{tabularx}
\end{table}

We record report identity, field parsing, and execution status separately for
each collection. Explicit no-alert outcomes remain distinct from missing
reports and execution failures. Findings already produced are retained
alongside execution status. We extract native category or rule identifiers,
descriptions, evidence locations, severity, scores, and native assessments.
We associate findings with the taxonomy while retaining the links between
each Skill, scanner, report, and finding.

For the threat-characteristic analysis of malicious Skills, we additionally
examine report text and Skill content to resolve undetermined facets.
This covers undetermined facets in mapped findings and empty facets in
findings whose category role remains unresolved. Non-threat items remain
excluded. We use \texttt{gpt-6-luna} to assist coding and produce candidates
supported by source quotations. Programmatic checks verify quotation
locations, finding identities, and item coverage. Model-based semantic checks
also examine non-supply-chain sources and target additions.
Candidates retain their rationales and applicability conditions.
These instance-level additions have not yet completed human review.

Each undetermined facet of a finding can receive at most one supported
additional value. Existing determined values remain, and different findings
may contribute different values. Insufficient or ambiguous evidence remains
undetermined. Scanner report comparisons use native category mappings,
whereas malicious-Skill characterization uses the context-informed instance
records. Sampling, identity matching, scanner versions and commands, field
rules, and the evidence protocol are documented in the supplementary
materials described in Appendix~\ref{app:platform-collection}.

\subsection{Analysis Methods and Metrics}
\label{sec:alignment-method}

\paragraph{Category relations and label representation.}
Drawing on ontology matching and the Simple Knowledge Organization System
(SKOS)~\cite{euzenat2013matching,skos2009}, we compare the scope of native
category definitions. We distinguish equivalence, containment, partial overlap,
and disjointness. We record the direction of containment
and mark cases with insufficient evidence as unresolved.
Appendix~\ref{app:taxonomy-relations} defines the decision rules.

For RQ1, we examine 90 category pairs, with 30 pairs each from comparisons
between benchmarks, between benchmarks and scanners, and between scanners.
Within each group, we sample pairs with different numbers of shared facet
values under the existing mappings. Two authors independently judge the
relations, and a third party adjudicates disagreements.

We also examine which facets benchmark labels express within their native
layers. We mark a label \emph{Full} if its definition expresses all three
facets and \emph{Partial} if it expresses only some.
Appendix~\ref{app:full-statistics} reports the results by layer.

\paragraph{Scanner report comparison.}
For RQ2, let $K_i(s)$ be the set of determined Mechanism values reported by scanner $i$
for Skill $s$, obtained through native category mappings and deduplicated
within that Skill. We compare two reports using Jaccard similarity,
\begin{equation}
  J_{ij}(s)=\frac{|K_i(s)\cap K_j(s)|}{|K_i(s)\cup K_j(s)|}.
  \label{eq:report-jaccard}
\end{equation}
The numerator counts shared Mechanisms and the denominator counts the union
of Mechanisms from the two scanners. We calculate this value when at least one
scanner has a determined Mechanism and average equally across Skills.
Pairs with two empty sets are recorded separately.
We also distinguish Skills with values from only one scanner, two nonempty but
disjoint sets, partially shared sets, and identical nonempty sets.
An empty set means that no determined Mechanism was obtained through the
category mapping. Findings and execution status remain separately recorded.

For each Skill, we count each Mechanism once across all scanner reports.
The same Mechanism in two different Skills counts as two units.
Let $U$ be the set of distinct (Skill, Mechanism) pairs and $U_1$ the
subset reported by exactly one scanner. The share of pairs reported by only
one scanner is
\begin{equation}
  P_{\mathrm{single}}=\frac{|U_1|}{|U|}.
  \label{eq:single-tool-proportion}
\end{equation}
For a scanner combination $B$, let $U_B$ contain the units supplied by that
combination. Its share of all units reported by the study scanners is
\begin{equation}
  C(B)=\frac{|U_B|}{|U|}.
  \label{eq:combination-proportion}
\end{equation}
Both denominators count Mechanisms deduplicated within each Skill and summed
across Skills. The two measures describe report complementarity and the
reported information retained by a scanner combination, respectively.
We do not calculate a proportion for an empty denominator and report the
eligible sample count for each result. Full combination and report-status
records accompany the supplementary results in
Appendix~\ref{app:full-statistics}.

\paragraph{Skill-level aggregation.}
For RQ3, we aggregate context-informed instance records by Skill.
Let $F(s)$ be the included findings from all scanners for Skill $s$, and let
$A_d(f)$ be the determined values of finding $f$ for facet
$d\in\{S,M,T\}$. We aggregate by Skill as
\begin{equation}
  V_d(s)=\bigcup_{f\in F(s)}A_d(f).
  \label{eq:skill-facet-values}
\end{equation}
A category count is the number of Skills containing that value, with all
317 Skills as the denominator for proportions. The quantity $|V_d(s)|$ is
the number of distinct values in the facet, and zero means that no value is
determined. A Skill may have multiple categories, so proportions may sum to
more than 100\%.
Pairwise or three-facet co-occurrence means that the values appear in the
same Skill, counting each combination once within it.
Exact-profile frequencies require all Source, Mechanism, and Target values
to match, retaining empty facets. Combinations directly supported within a
single finding are recorded separately. Full distributions and combinations
are documented in the supplementary results described in
Appendix~\ref{app:full-statistics}.

%% file: sections/08-results-rq1.tex
\section{Results}
\label{sec:results}

\subsection{RQ1: Threat Information and Relations in Native Categories}
\label{sec:rq1-results}

Native categories emphasize different aspects of a threat. For example, the
name \emph{Supply Chain Attack} identifies a supply-chain concern without
specifying the operation or attack objective. Cisco's definition further
describes malicious package or tool injection, supporting a supply-chain
Source and dependency-manipulation Mechanism, while leaving the Target
undetermined~\cite{ciscotaxonomy}. Understanding what an attack does therefore
requires examining definitions and report evidence as well as category names.

Among the 168 threat categories in the inventory drawn from four scanners,
one security standard, and one threat rule system, four have undetermined
Sources (2.4\%), 11 have undetermined Mechanisms (6.5\%), and 99 have
undetermined Targets (58.9\%). Another 105 categories
(62.5\%) are recorded as source-independent, meaning that their definitions
do not restrict the source. This differs from the four categories whose
Sources cannot be determined. Targets are most often unspecified.
Many categories describe an operation or risk condition without identifying
the consequence it pursues. Appendix~\ref{app:full-statistics} reports
benchmark label representation within each native layer.

\finding{Some native risk categories do not jointly describe threat source,
mechanism, and target. Among the 168 included threat categories, 58.9\% leave
the Target undetermined. Interpreting a particular Skill therefore requires
instance evidence beyond the category definition.}

Category names also differ from the scope of their definitions.
Caterpillar's \emph{Persistence mechanisms} and MalSkillBench's
\emph{B5 Persistence} have different names but both cover cron, startup
scripts, or related persistence mechanisms~\cite{caterpillardocs,malskillbench2026}.
We judge these categories equivalent.
Conversely, similar names do not guarantee equal scope.
Caterpillar's \emph{Data exfiltration} focuses on outbound transfer using
curl or wget, whereas SkillSpector's \emph{Data Exfiltration (E1--E4)} also
includes secret collection, sensitive-directory inspection, and session-context
transfer~\cite{caterpillardocs,skillspectordocs}. The latter contains the
former. Cross-source alignment therefore
requires comparing the operations, objects, and applicability conditions in
the definitions.

Table~\ref{tab:native-category-relations} summarizes the 90 sampled category
pairs, with 30 from each group. These proportions describe the sampled pairs. There are 26 partial overlaps, nine
containment relations, and one equivalent relation, totaling 36 pairs.
The remaining 54 are disjoint. For example, \emph{B5 Persistence} and
SkillTrustBench's \emph{Agent Memory Poisoning} describe system-level
persistence and persistent-memory content poisoning, respectively, and are
judged disjoint. Both can occur in a Skill without their category definitions
overlapping. The category-relation records retain the compared definitions,
rationales, and containment directions, as described in
Appendix~\ref{app:taxonomy-relations}.

\begin{table}[t]
  \caption{Relations among the 90 selected native category pairs. The sample
  includes comparisons between benchmarks, between benchmarks and scanners,
  and between scanners.}
  \label{tab:native-category-relations}
  \centering
  \small
  \begin{tabular}{lrr}
    \toprule
    Relation & Pairs & Share \\
    \midrule
    Equivalent & 1 & 1.1\% \\
    Partial overlap & 26 & 28.9\% \\
    Containment & 9 & 10.0\% \\
    Disjoint & 54 & 60.0\% \\
    \bottomrule
  \end{tabular}
\end{table}

\finding{Among the 90 selected native category pairs, 40\% partially overlap,
contain one another, or are equivalent in definition. Different risk names
can describe shared threat content, so comparison across classification
systems requires definition-level correspondences.}

%% file: sections/09-results-rq2.tex
\subsection{RQ2: Shared and Different Mechanisms in Scanner Reports}
\label{sec:rq2-results}

We compare scanner reports for 386 marketplace Skills and 317 malicious Skills
from MalSkillBench. Table~\ref{tab:report-jaccard} reports the mean Jaccard
similarity between Mechanism sets obtained through native category mappings,
with equal weight for each eligible Skill. Means range from 0.000 to 0.169
for marketplace scanner pairs and from 0.028 to 0.171 for malicious-Skill
scanner pairs. Different scanners frequently provide different Mechanism descriptions
for the same Skill.

\begin{table}[tbp]
  \caption{Jaccard similarity of reported Mechanism sets.}
  \label{tab:report-jaccard}
  \centering
  \small
  \setlength{\tabcolsep}{3pt}
  \begin{tabularx}{\columnwidth}{@{}>{\raggedright\arraybackslash}Xrrr@{}}
    \toprule
    Scanner pair & $n_+$ & $n_0$ & $\bar J$ \\
    \midrule
    \multicolumn{4}{@{}l}{\emph{Marketplace reports}} \\
    SkillSpector / Snyk & 206 & 180 & 0.059 \\
    SkillSpector / Socket & 115 & 271 & 0.007 \\
    SkillSpector / Agent Trust Hub & 271 & 115 & 0.052 \\
    Snyk / Socket & 152 & 234 & 0.000 \\
    Snyk / Agent Trust Hub & 274 & 112 & 0.169 \\
    Socket / Agent Trust Hub & 252 & 134 & 0.001 \\
    \midrule
    \multicolumn{4}{@{}l}{\emph{Malicious-Skill reports}} \\
    Cisco / SkillSpector & 290 & 27 & 0.028 \\
    Cisco / Caterpillar & 308 & 9 & 0.171 \\
    SkillSpector / Caterpillar & 237 & 80 & 0.056 \\
    \bottomrule
  \end{tabularx}
  \par\smallskip
  \begin{minipage}{\columnwidth}
    \footnotesize\raggedright
    $n_+$ counts Skills with determined Mechanisms from at least one scanner;
    $n_0$ counts Skills with two empty sets. $\bar J$ is the mean Jaccard
    similarity over the $n_+$ Skills.
  \end{minipage}
\end{table}

A low mean similarity can arise when only one scanner reports a Mechanism
or when both report different Mechanisms. For SkillSpector and Socket, at
least one scanner reports a Mechanism for 115 marketplace Skills. Of these,
112 have values only from SkillSpector, one only from Socket, and two have
partly overlapping sets. For SkillSpector and Snyk, 206 Skills meet the same
criterion. Of these, 55 have values only from SkillSpector, 92 only from Snyk,
37 have disjoint nonempty sets, 17 have partly overlapping sets, and five
have identical sets.

Cisco and Caterpillar reports for two MalSkillBench samples illustrate
these differences. For \emph{metamask},
Cisco categorizes a remote installation command as \emph{Supply Chain Attack},
emphasizing unpinned dependency installation. Caterpillar's
\emph{Data Exfiltration} finding describes the same
\texttt{curl ... | bash} command as remote code execution.
The findings map to dependency manipulation and code execution, respectively.
Their Mechanism sets are disjoint even though the findings refer to the same
command.

Both scanners report package-installation risks in a second sample
presented as a browser automation Skill called \emph{Agent Browser}.%
\footnote{MalSkillBench sample ID: \texttt{agent-browser-b2x7tvcmbjgp}.}
Cisco also flags a mismatch between the Skill's declared Bash permissions
and the commands in its instructions.

\finding{Scanners frequently describe different Mechanisms for the same Skill.
These differences include additional reported risks and different
classifications of the same operation. Evidence in the original reports helps
distinguish the two.}

Counting each (Skill, Mechanism) pair once across scanner reports yields
773 pairs for marketplace Skills and 828 for malicious Skills. Of these,
618 marketplace pairs (79.9\%) and 682 malicious-Skill pairs (82.4\%) appear
in only one scanner's reports. In the marketplace collection, 418 pairs
appear only in Agent Trust Hub reports and 134 only in SkillSpector reports.
In the malicious collection, 461 appear only in Cisco reports and 173 only in
Caterpillar reports.

Table~\ref{tab:tool-combinations} lists representative combinations with the
most reported Mechanism units of each size. Agent Trust Hub alone
provides 72.3\% of marketplace units, increasing to 91.5\% when combined with
SkillSpector. Adding Snyk matches the union of all four marketplace scanners.
For malicious Skills, Cisco alone provides 70.3\%, and combining it with
Caterpillar reaches 94.2\%. Adding SkillSpector includes all reported units.
Under the current category mappings, Socket adds no Mechanism unit absent
from the other marketplace scanners.

\begin{table}[tbp]
  \caption{Mechanism units retained by scanner combinations.}
  \label{tab:tool-combinations}
  \centering
  \small
  \setlength{\tabcolsep}{3pt}
  \begin{tabularx}{\columnwidth}{@{}>{\raggedright\arraybackslash}Xrr@{}}
    \toprule
    Scanner or combination & Count & Share \\
    \midrule
    \multicolumn{3}{@{}l}{\emph{Marketplace reports}} \\
    Agent Trust Hub & 559 & 72.3\% \\
    SkillSpector + Agent Trust Hub & 707 & 91.5\% \\
    SkillSpector + Snyk + Agent Trust Hub & 773 & 100.0\% \\
    \midrule
    \multicolumn{3}{@{}l}{\emph{Malicious-Skill reports}} \\
    Cisco & 582 & 70.3\% \\
    Cisco + Caterpillar & 780 & 94.2\% \\
    Cisco + SkillSpector + Caterpillar & 828 & 100.0\% \\
    \bottomrule
  \end{tabularx}
  \par\smallskip
  \begin{minipage}{\columnwidth}
    \footnotesize\raggedright
    Mechanisms are deduplicated within each Skill and summed across Skills.
    Shares use the unions reported by all study scanners as denominators,
    comprising 773 marketplace units and 828 malicious-Skill units.
  \end{minipage}
\end{table}

\finding{Of the distinct (Skill, Mechanism) pairs, 79.9\% in the marketplace
cohort and 82.4\% in the malicious-Skill cohort appear in only one scanner's
reports. The selected two-scanner combinations retain 91.5\% and 94.2\% of
all pairs reported for the respective cohorts.}

%% file: sections/10-results-rq3.tex
\subsection{RQ3: Sources, Mechanisms, and Targets in Malicious Skills}
\label{sec:rq3-results}

We characterize the 317 malicious Skills selected from MalSkillBench using
the context-informed instance records described in Section~\ref{sec:corpora}.
For each Skill, we combine findings across scanners and deduplicate their values.
Category counts use Skills as the unit, with 317 as the denominator for all
proportions.

Table~\ref{tab:malicious-skill-categories} shows that supply chain is the most
frequent Source (97.2\%), while some Skills also involve external untrusted
content or user input. Dependency manipulation, code execution, and
obfuscation or hiding are the most frequent Mechanisms. Defense evasion and
information theft are the most frequent Targets. Repeated combinations also
occur. Supply-chain Source, dependency-manipulation Mechanism, and
defense-evasion Target co-occur in 185 Skills (58.4\%), the most frequent
three-facet combination. Dependency manipulation and code execution co-occur
in 169 Skills (53.3\%), while defense evasion and information theft co-occur
in 73 (23.0\%).

\begin{table}[tbp]
  \caption{Category counts and proportions for malicious Skills.}
  \label{tab:malicious-skill-categories}
  \centering
  \small
  \setlength{\tabcolsep}{3pt}
  \begin{tabularx}{\columnwidth}{@{}>{\raggedright\arraybackslash}Xr@{}}
    \toprule
    Facet value & $n$ (\%) \\
    \midrule
    \multicolumn{2}{@{}l}{\emph{Source}} \\
    Supply chain & 308 (97.2\%) \\
    External untrusted content & 80 (25.2\%) \\
    User input & 19 (6.0\%) \\
    Runtime environment & 1 (0.3\%) \\
    No determined value & 9 (2.8\%) \\
    \midrule
    \multicolumn{2}{@{}l}{\emph{Mechanism}} \\
    Dependency manipulation & 261 (82.3\%) \\
    Code execution & 196 (61.8\%) \\
    Obfuscation / hiding & 159 (50.2\%) \\
    Instruction manipulation & 108 (34.1\%) \\
    Trigger abuse & 101 (31.9\%) \\
    Privilege abuse & 69 (21.8\%) \\
    State corruption & 26 (8.2\%) \\
    Adaptive evasion & 6 (1.9\%) \\
    Subagent privilege gain & 0 (0.0\%) \\
    No determined value & 9 (2.8\%) \\
    \midrule
    \multicolumn{2}{@{}l}{\emph{Target}} \\
    Defense evasion & 224 (70.7\%) \\
    Information theft & 120 (37.9\%) \\
    System damage & 87 (27.4\%) \\
    Content-safety harm & 81 (25.6\%) \\
    Persistent control & 26 (8.2\%) \\
    Resource abuse & 18 (5.7\%) \\
    Financial theft & 0 (0.0\%) \\
    No determined value & 18 (5.7\%) \\
    \bottomrule
  \end{tabularx}
  \par\smallskip
  \begin{minipage}{\columnwidth}
    \footnotesize\raggedright
    Each facet is sorted independently. Skills may have multiple values.
    Skills without a determined value are listed separately.
  \end{minipage}
\end{table}

\begin{figure*}[t]
  \centering
  \includegraphics[width=\textwidth]{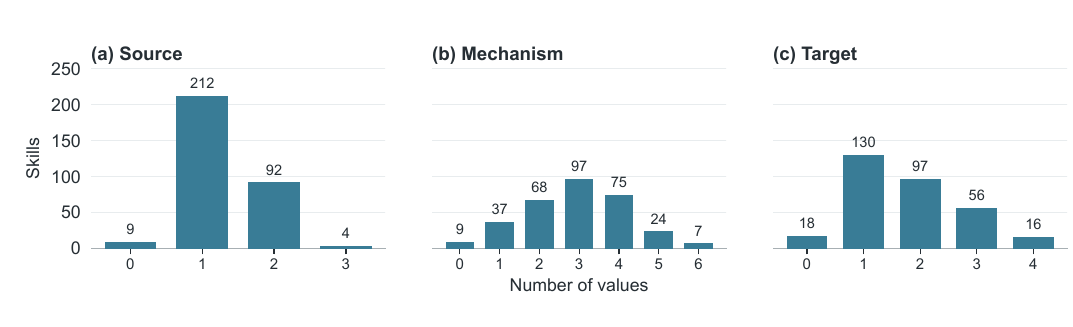}
  \caption{Numbers of distinct Source, Mechanism, and Target values per Skill.
  The horizontal axes give value counts and the vertical axes give Skill
  counts. All three panels use the same vertical scale and contain 317 Skills.
  Zero denotes the absence of a determined value in that facet.}
  \Description{Three side-by-side distributions show the number of distinct
  sources, mechanisms, and targets associated with each malicious Skill.
  Mechanisms and targets frequently have multiple values.}
  \label{fig:skill-facet-cardinality}
\end{figure*}

Figure~\ref{fig:skill-facet-cardinality} further shows that 271 Skills
(85.5\%) have at least two Mechanisms and 169 (53.3\%) have at least two
Targets. There are 96 Skills (30.3\%) with at least two Sources.
The median number of Mechanisms is three, with a maximum of six.
The median number of Targets is two, with a maximum of four.
In total, 295 Skills (93.1\%) have determined values in all three facets.
Common risk objectives therefore often co-occur with multiple operating
Mechanisms within the same Skill.

\finding{Malicious Skills commonly have multiple Mechanisms and Targets.
At least two Mechanisms appear in 85.5\% of Skills, and at least two Targets
appear in 53.3\%. A single risk category is insufficient to summarize their
full threat characteristics.}

The \emph{botroast} case illustrates behavioral differences within a single
native category. Cisco places findings about periodic submission, expanded
data collection, and transmission to an external API under \emph{Data Exfiltration}. This category names the
information-theft objective without distinguishing the operating Mechanisms
in its label. Using report descriptions and Skill content,
Table~\ref{tab:botroast-mechanisms} distinguishes trigger abuse, privilege
abuse, and code execution. All three concern collection and submission
instructions distributed with the Skill and share an information-theft
Target. Periodic triggering applies only when the corresponding heartbeat is
enabled.

\begin{table}[tbp]
  \caption{Different Mechanisms within the same native category in
  \emph{botroast}.}
  \label{tab:botroast-mechanisms}
  \centering
  \small
  \begin{tabularx}{\columnwidth}{@{}>{\raggedright\arraybackslash}X@{\hspace{7pt}}>{\raggedright\arraybackslash}p{0.26\columnwidth}@{}}
    \toprule
    Report description and supporting content & Mechanism \\
    \midrule
    Collect and submit data every 24 hours when the corresponding heartbeat is enabled & Trigger abuse \\
    Expand collection to arbitrary content in the workspace & Privilege abuse \\
    Read personal files and submit them to an external API using curl & Code execution \\
    \bottomrule
  \end{tabularx}
  \par\smallskip
  \begin{minipage}{\columnwidth}
    \footnotesize\raggedright
    All three findings have the native category \emph{Data Exfiltration}
    and the Target \emph{Information theft}. Facet values are assigned from
    the reports and Skill content.
  \end{minipage}
\end{table}

We count the shared information-theft Target once for this Skill and retain
the different Mechanisms. This representation
distinguishes periodic triggering, expanded access, and data submission under
a shared objective, rather than counting only multiple identically named
data-exfiltration alerts. Exact-profile frequencies and item-level evidence
are retained in the supplementary materials described in
Appendices~\ref{app:full-statistics} and~\ref{app:platform-collection}.

\finding{A single native risk category can contain different Mechanisms.
Analysis of Source, Mechanism, and Target using report and Skill content
distinguishes these Mechanisms and their shared objective, exposing threat
characteristics that a single label leaves unspecified.}

%% file: sections/11-discussion.tex
\section{Discussion and Software Engineering Implications}
\label{sec:discussion}

Our study reveals differences in scanner categories and reports and shows
how semantic alignment helps explain shared risks and additional findings.
Based on these results, we call for action by scanner vendors, Skill
platforms and users, and researchers.

\subsection{A Shared Risk Taxonomy for Scanner Vendors}

Scanners maintain their own risk categories, making it difficult for
developers and users to understand the relationships between reports.
We call on scanner vendors to jointly establish and maintain a shared Skill
risk taxonomy. Consistent categories can describe the Source, Mechanism,
and Target of a risk, with explicit mappings from native categories to the
shared taxonomy.

A common classification basis can help developers and users understand the
malicious behaviors described in reports. It can also help vendors compare
their detection scopes and identify overlapping or uncovered risks.
Retaining native reports and behavioral evidence while aligning and
integrating findings can support collaboration among scanners and help
improve the identification and detection of malicious Skills.

\subsection{Interpreting Multi-Scanner Consensus}

When multiple scanners judge the same Skill to be low risk, their judgments
do not necessarily provide independent security evidence.
If scanners focus on largely overlapping Mechanisms, they may also miss similar
threats. Agreement on low risk may therefore coexist with shared blind
spots. The number of agreeing scanners alone cannot establish a Skill's safety.

RQ2 shows both intersections between reports and additional information
provided by individual scanners. We call on Skill platforms and users to
consider both overlap and complementarity when interpreting multiple reports
and to examine what different risk information an additional scanner provides.
Presenting shared findings and additional findings separately can help users
understand the security evidence provided by multiple reports and plan
further review.

\subsection{Investigating Disagreement and Combination Benefits}

The cases in RQ2 show that scanners can classify the same operation differently
or report additional behaviors. We call on researchers to investigate the
causes of these differences and the conditions under which complementary
information leads to actual detection gains.

One direction is to construct paired samples that preserve the same risk
behavior while changing only the instruction wording or implementation
form. With scanner versions and analysis conditions held constant, researchers
can examine changes in findings and decisions.
Comparing these changes with behavioral evidence can reveal sensitivity to
expression and distinguish changes in classification from differences in
detection. Another direction is to use independently verified risk behaviors
to evaluate how scanner combinations cover multiple Mechanisms within the same
Skill, measuring additional correct detections and threats missed by all
scanners.

%% file: sections/12-threats-ethics.tex
\section{Threats to Validity and Ethics}
\label{sec:threats}

\paragraph{Taxonomy completeness.}
We construct the taxonomy by drawing on existing approaches to threat
classification and analyzing scanner categories, academic papers, and industry
materials. Although these materials cover a range of risks and their
manifestations, the taxonomy may omit threats that the sources have not
recorded or adequately described. As the Skill ecosystem and attack methods
evolve, its categories and boundaries may require further refinement and
extension.

\paragraph{Subjectivity in manual coding.}
Taxonomy construction involves interpreting threat descriptions and assigning
categories. The authors' knowledge and experience may therefore affect coding.
To reduce individual bias, two authors independently coded the material for
taxonomy construction, and a third party adjudicated disagreements.
This process helps improve consistency, although shared omissions or
interpretive biases may remain.

\paragraph{Selection bias in marketplace samples.}
The marketplace samples come from publicly listed Skills and their security
reports. A platform may block a malicious Skill before publication or remove
it once malicious behavior is identified. These Skills would then be unavailable
for collection. Platform screening may therefore shape the risk composition
of publicly visible samples. Scanner differences observed on these samples
may not fully reflect how the scanners behave on malicious Skills.
We additionally scan malicious samples from MalSkillBench to supplement
malicious content potentially missing from the marketplace data.
These two data sources still provide limited coverage of the broader Skill
ecosystem.

\paragraph{Research ethics.}
This study uses public marketplace reports and existing benchmark material
to support the identification of malicious Skills and defenses against them.
We treat Skill contents as untrusted static material and did not install
malicious Skills or execute their malicious workflows.
To reduce potential misuse of sensitive information, access tokens in the
paper's examples are displayed as placeholders.

%% file: sections/13-conclusion.tex
\section{Conclusion}
\label{sec:conclusion}

We develop a unified Skill risk taxonomy with Source, Mechanism, and Target
facets and use it to analyze existing risk classifications, scanner reports,
and the threat characteristics of malicious Skills.
Our study reveals incomplete threat descriptions and overlapping semantic
boundaries in native categories. Reports for the same Skill also differ
substantially. Semantic alignment reveals shared Mechanisms and additional
findings within these reports.
Of the distinct (Skill, Mechanism) pairs, 79.9\% for marketplace Skills and
82.4\% for malicious Skills appear in only one scanner's reports. Malicious Skills also commonly involve
multiple Mechanisms and Targets, making a single risk label insufficient to
describe their threats fully.

These results show that a common classification basis helps explain scanner
differences and integrate findings across scanners. We call on scanner vendors
and Skill platforms to jointly establish and maintain a shared risk taxonomy
and standardized reporting to support scanner improvement, evaluation, and
selection. Future research can investigate the causes of scanner
disagreement and evaluate the actual detection gains from combining scanners.

%% file: sections/15-taxonomy-relations-appendix.tex
\section{Taxonomy and Category-Relation Definitions}
\label{app:taxonomy-relations}

\subsection{Facet Boundaries and Mapping Records}

Source identifies the trust origin or input boundary through which malicious
influence enters. It is distinct from the file or medium carrying a payload.
Mechanism identifies the operating process, including changes to instructions,
execution, dependencies, permissions, or state. Target identifies an
explicitly pursued or caused security consequence. Weakness, carrier,
lifecycle, severity, and product disposition are retained separately.

A definition may support multiple values within a facet. An unexpressed facet
remains undetermined. Source-independent records a definition that explicitly
does not constrain the origin, whereas undetermined records insufficient
source information. Non-threat controls are kept separate from threat
categories. Instance-level additions based on report and Skill content retain
their own evidence and do not overwrite a native category definition.

Detailed category records retain the native identifier, source, definition,
hierarchical position, facet assignments, evidence location, and mapping
rationale. Benchmark label layers and scanner identifiers retain their native
granularity. The records distinguish a category from a particular finding
reported for a Skill. The source documentation provides the original category
boundaries~\cite{ciscotaxonomy,snykissuecodes,socketalerttypes,skillspectordocs,aartsstandard,atrdocs,caterpillardocs}.
Benchmark definitions are retained with their respective label
systems~\cite{malskillbench2026,maliciousskillbench2026,skilltrustbench2026}.

\subsection{Category-Relation Decisions}

We compare native definitions at the level of the same security instance.
Equivalent definitions admit the same instances within the comparison scope.
Containment requires every instance of the narrower definition to satisfy
the broader definition, with additional instances admitted by the broader
definition. Partial overlap requires at least one shared instance and
instances exclusive to each definition. Disjoint definitions admit no shared
instance. Insufficient or conflicting evidence is recorded as unresolved.

Containment direction is retained in the category-pair record even when the
results aggregate both directions as containment. A shared facet value is
not sufficient to establish equivalence. A causal connection, detection
relation, prerequisite, or possible co-occurrence does not by itself establish
that two definitions describe the same semantic object.

For example, persistence through system services and poisoning of persistent
agent memory can co-occur in a Skill while denoting different security
instances. Conversely, differently named definitions can cover the same
persistence behavior. Each comparison retains both native definitions, its
relation and direction where applicable, and the supporting rationale in the
accompanying category-relation records.

%% file: sections/16-platform-collection-appendix.tex
\section{Marketplace Collection, Matching, and Report Inclusion}
\label{app:platform-collection}

\subsection{Cross-Platform Candidate Discovery and Content Checks}

Collection targets Skills listed on both skills.sh and ClawHub. Candidate
scripts associate names with authors, repositories, and platform URLs.
When a name has several possible authors or repositories, the scripts expand
the possible links and check collected \texttt{SKILL.md} content.
The 1,082 records represent cross-platform candidates produced by these
scripts. The 386 compared Skills are the subset retained after requiring
reports from all four study scanners.

Content comparison normalizes whitespace before computing word-set Jaccard
similarity over the first 200 characters and character 20-gram overlap over
the whole document. The verification and merging step requires at least 80\%
for the former and 50\% for the latter. The following excerpt preserves the
verification predicate from \texttt{verify\_and\_merge.py}, with a diagnostic
print omitted. Candidate and verification records retain platform URLs and
identity information.

\begingroup
\footnotesize
\begin{verbatim}
def passes_verification(item):
    v = get_verification(
        item['owner'], item['repo'],
        item['skill_slug'])
    if not v:
        return False
    p200 = v.get('prefix200_jaccard_pct', 0)
    ngram = v.get('ngram_overlap_pct', 0)
    return p200 >= 80 and ngram >= 50
\end{verbatim}
\endgroup

\subsection{Report Collection and Four-Scanner Inclusion}

Collection scripts save public audit pages and linked reports and associate
each Skill, platform URL, scanner, and raw file. Not every platform Skill has
a public result from every scanner. A candidate is included in the four-scanner
comparison only when SkillSpector, Snyk, Socket, and Agent Trust Hub all have
traceable, readable, and parseable report sources. An explicit no-alert result
satisfies report availability. Missing reports, parsing failures, and
execution status remain separately recorded.

The report-comparison implementation forms the common cohort using each
scanner's \texttt{available} status. Its inclusion expression is
\begingroup
\footnotesize
\begin{verbatim}
projection = [
    s for s in skills
    if all(records[(s, t)]['available']
           for t in tools)
]
\end{verbatim}
\endgroup
Here, \texttt{tools} contains the four marketplace scanners.
Availability checks whether retained report records and result fields can be
read and parsed, rather than whether they contain an alert. This rule yields
386 Skills for comparison, and scanner differences are calculated on that fixed
cohort.

\subsection{Collection and Analysis Records}

\setcounter{table}{0}
\renewcommand{\thetable}{B\arabic{table}}
\begin{table}[tbp]
  \caption{Scripts associated with marketplace discovery and report inclusion.}
  \label{tab:collection-scripts}
  \centering
  \small
  \renewcommand{\arraystretch}{1.05}
  \begin{tabularx}{\columnwidth}{@{}>{\raggedright\arraybackslash}p{0.47\columnwidth}@{\hspace{7pt}}>{\raggedright\arraybackslash}X@{}}
    \toprule
    Step and script & Purpose \\
    \midrule
    \textbf{Candidate discovery}\newline \texttt{cartesian\_\allowbreak match.py} & Find matching names and expand possible author and repository links. \\
    \textbf{Content collection}\newline \texttt{fetch\_\allowbreak skill\_\allowbreak md.py} & Retrieve Skill content and calculate prefix and whole-document overlap. \\
    \textbf{Verification and merging}\newline \texttt{verify\_\allowbreak and\_\allowbreak merge.py} & Apply content thresholds and merge candidate records. \\
    \textbf{Report collection}\newline \texttt{fetch.py} & Save public pages, reports, and collection status. \\
    \textbf{Report inclusion and comparison}\newline \texttt{analyze\_\allowbreak existing\_\allowbreak reports\_\allowbreak v5.py} and its base implementation & Read retained evidence, construct fixed common cohorts, and compare reports. \\
    \bottomrule
  \end{tabularx}
\end{table}

Table~\ref{tab:collection-scripts} identifies the scripts for the marketplace
pipeline. The accompanying records retain candidate lists, content checks,
report indexes, and inclusion lists. Scanner versions and commands, field
rules, sampling records, and the complete evidence-checking protocol are
retained with the study materials. The main text summarizes collection
targets, collection and cohort sizes, and principal inclusion rules.

%% file: sections/17-full-statistics-appendix.tex
\section{Supplementary Statistics and Label Representation}
\label{app:full-statistics}
\suppressfloats[t]

\subsection{Benchmark Label Representation}

Table~\ref{tab:benchmark-label-representation} reports Full and Partial labels
within their native layers, using the definitions in
Section~\ref{sec:alignment-method}. Each row has its own label denominator.
Components from different layers are not pooled into a mutually exclusive
category population.

\setcounter{table}{0}
\renewcommand{\thetable}{C\arabic{table}}
\begin{table}[t]
  \caption{Representation of the three facets in native benchmark label layers.}
  \label{tab:benchmark-label-representation}
  \centering
  \small
  \setlength{\tabcolsep}{3pt}
  \begin{tabularx}{\columnwidth}{Xrrr}
    \toprule
    Native label layer & Labels & Full & Partial \\
    \midrule
    MalSkillBench components & 28 & 8 & 20 \\
    SkillTrustBench threats & 8 & 4 & 4 \\
    MaliciousSkillBench attacks & 11 & 2 & 9 \\
    MaliciousSkillBench impacts & 9 & 0 & 9 \\
    \bottomrule
  \end{tabularx}
\end{table}

\subsection{Category Inventory and Exact Profiles}

Taxonomy category-representation statistics use the 172 mappings from six
direct sources. Of these, 168 are threat categories and four are controls.
Controls do not enter the threat-category denominator. Native definitions and
mapping rationales are retained in the category records described in
Appendix~\ref{app:taxonomy-relations}.

The 317 malicious Skills form 158 distinct Source, Mechanism, and Target
value profiles, retaining facets without a determined value. The most
frequent exact profile contains supply chain as Source, dependency
manipulation, code execution, and obfuscation or hiding as Mechanisms, and
defense evasion as Target. It occurs in 21 Skills (6.6\%). A second profile
that additionally contains trigger abuse occurs in 17 (5.4\%).
Considered separately, the facets form six distinct Source sets, 58 Mechanism
sets, and 28 Target sets. Full pairwise relations, scanner combinations,
co-occurrence matrices, and exact-profile frequencies remain in the
accompanying supplementary results.

\subsection{Complete Counting Definitions}

For a fixed Skill cohort $D$ and scanner set $\mathcal T$, let $F_i(s)$ contain
the included findings of scanner $i$ for Skill $s$, and let $M(f)$ contain the
determined Mechanisms obtained through the native category mapping of
finding $f$. Then
\[
  K_i(s)=\bigcup_{f\in F_i(s)}M(f).
\]
Finding roles and undetermined Mechanisms remain separately recorded.
We define the eligible set and equally weighted mean as
\[
  D_{ij}^{+}=\{s\in D:K_i(s)\cup K_j(s)\ne\varnothing\},
  \qquad
  \overline J_{ij}=\frac{1}{|D_{ij}^{+}|}
    \sum_{s\in D_{ij}^{+}}J_{ij}(s).
\]
Two empty sets do not receive a similarity of one and are counted separately.
No mean is calculated when the eligible set is empty. Both a single nonempty
side and two disjoint nonempty sets have similarity zero, but their relation
counts remain distinct.

For aggregate report comparison,
\[
  E_i=\{(s,m):s\in D,\ m\in K_i(s)\},
  \qquad U=\bigcup_{i\in\mathcal T}E_i.
\]
Each unit identifies a Skill and a Mechanism. Repetitions within one Skill
are deduplicated, whereas the same Mechanism in different Skills contributes
different units. Let $n(u)$ count the scanners reporting unit $u$. Then
\[
  U_1=\{u\in U:n(u)=1\},
  \qquad U_B=\bigcup_{i\in B}E_i,
  \quad B\subseteq\mathcal T.
\]
The measures in Equations~\ref{eq:single-tool-proportion}
and~\ref{eq:combination-proportion} use the same reported union $U$ as their
denominator. Units reported by several scanners are counted once in each union
and are excluded from $U_1$. Empty denominators remain uncalculated.

For the malicious-Skill cohort $D_{\mathrm{mal}}$, the count of a facet value
and its proportion are
\[
  N_d(v)=|\{s\in D_{\mathrm{mal}}:v\in V_d(s)\}|,
  \qquad
  p_d(v)=\frac{N_d(v)}{|D_{\mathrm{mal}}|}.
\]
Here, $|D_{\mathrm{mal}}|=317$. The denominator retains Skills without a
determined value in that facet. Unknown or applicability states are not
inserted as threat values, and non-threat items remain excluded.
The number of values is $k_d(s)=|V_d(s)|$.
Zero includes Skills without a determined value, and each facet's cardinality
distribution totals 317 Skills.

Co-occurrence of values $v_a$ and $v_b$ counts Skills satisfying both
membership conditions,
\[
  N_{ab}(v_a,v_b)=
  |\{s\in D_{\mathrm{mal}}:
      v_a\in V_a(s),\ v_b\in V_b(s)\}|.
\]
Three-facet co-occurrence adds a third membership condition. Co-occurrence
within a facet requires two different values to occur in its set.
Each combination counts once per Skill, with 317 as the denominator for
overall proportions. Values may be supplied by different findings.
Direct single-finding combinations are recorded separately.

Exact-profile frequencies group Skills only when their value-set tuples
\[
  (V_S(s),V_M(s),V_T(s))
\]
match completely, including empty sets.
Exact sets for one facet use the same rule on that facet alone.
Co-occurrence allows additional values, while exact-set frequencies place
Skills with additional values in a different group.

%% file: sections/14-appendix-corpora.tex
\section{Supplementary Historical Corpora}
\label{app:historical-corpora}

\paragraph{Generated-231.}
Each of the 231 generated entries has a matching artifact directory and a
\texttt{\_provenance.json} file recording its intended source, mechanism,
target, realization parameters, model, generation time, and file list.
Underscore-prefixed metadata was excluded from scanner input. Cisco and
Caterpillar each retain 231/231 raw reports and SkillSpector retains 227/231.
These reports enable finding-level supplementary analysis, but the malicious
label records construction intent rather than uniformly runtime-verified
behavior. We therefore do not use this corpus for strict true-positive or
false-negative claims.

\paragraph{Nominally benign-500.}
The frozen matrix contains aggregate fields from all three scanners for 500
samples, so the sample-by-scanner outcome matrix is complete. Item-level raw
findings are incomplete. The repository retains 13 Cisco, 10 SkillSpector, and 11
Caterpillar reports. Git history confirms that the aggregate rows were added in
two batches, but does not recover a uniform run manifest or the remaining raw
evidence. Moreover, membership in an upstream benign directory does not
establish verified benign status. Legitimate but security-sensitive
capabilities may reasonably trigger a scanner. We therefore call these samples
nominally benign references and use them only for exploratory aggregate profiles, not
formal false-positive rate or precision.

The historical Caterpillar documentation describes the offline scanner and
its grade/score output~\cite{caterpillardocs}. This provides product context.
The retained reports and study notes, rather than documentation alone,
establish the observed outcomes and available run provenance.

%% file: sections/18-taxonomy-sources-appendix.tex
\clearpage
\section{Academic and Industry Sources for Taxonomy Construction}
\label{app:taxonomy-sources}
\setcounter{table}{0}
\renewcommand{\thetable}{E\arabic{table}}

Tables~\ref{tab:academic-sources-a} and~\ref{tab:academic-sources-b} list
the academic sources consulted during taxonomy construction.
They include threat analyses, surveys, technical reports, and benchmarks.
Table~\ref{tab:industry-sources} lists the industry reports, standards,
frameworks, and practitioner guidance.
Each entry identifies a source and the classification perspective examined.
The associated references give document titles and source locators.

The sources address Skills and the wider agent ecosystem, including the
Model Context Protocol (MCP).
We use their threat descriptions, category definitions, and classification
criteria to develop the facets and values described in
Section~\ref{sec:taxonomy-method}.
Governance controls and risk tiers provide contextual classification
perspectives and remain distinct from attack categories.
Scanner definitions and benchmark labels used in the individual analyses
are identified in Section~\ref{sec:overview} and
Appendix~\ref{app:taxonomy-relations}.

\begin{table}[htbp]
  \caption{Academic sources for taxonomy construction, first group.}
  \label{tab:academic-sources-a}
  \centering
  \small
  \renewcommand{\arraystretch}{1.10}
  \begin{tabularx}{\columnwidth}{@{}>{\raggedright\arraybackslash}X@{\hspace{7pt}}>{\raggedright\arraybackslash}X@{}}
    \toprule
    Source & Classification perspective \\
    \midrule
    Coding-assistant attacks~\cite{maloyan2026coding} & Delivery vectors and attack modalities \\
    Skill-Inject~\cite{skillinject2026} & Skill-file injection and attacker control \\
    Agentic supply-chain runtime~\cite{taxonomy260219555} & Data and tool supply-chain stages \\
    MalSkillBench~\cite{malskillbench2026} & Attack vectors, behaviors, and insertion \\
    Skill supply-chain poisoning~\cite{taxonomy260403081} & Payload execution through documentation \\
    Taming OpenClaw~\cite{taxonomy260311619} & Threats across the agent lifecycle \\
    OpenClaw analysis and defense~\cite{taxonomy260310387} & Adversarial scenarios and security boundaries \\
    Malice in Agentland~\cite{taxonomy251005159} & Agent supply-chain backdoor threat models \\
    SkillFortify~\cite{taxonomy260300195} & Skill supply-chain threats and capabilities \\
    Coding assistants as an attacker's shell~\cite{taxonomy260525871} & External-artifact attack surfaces \\
    OpenClaw framework security~\cite{taxonomy260327517} & Architectural layers and trust violations \\
    SkillProbe~\cite{skillprobe2026} & Behavioral inconsistency and composition \\
    VIGIL~\cite{taxonomy260626524} & Behavioral policies and runtime violations \\
    Breaking the Protocol~\cite{taxonomy260117549} & Protocol trust and capability boundaries \\
    Prompt injections to protocol exploits~\cite{taxonomy250623260} & Input, system, and protocol threats \\
    Layered attack surface survey~\cite{chu2026lasm} & Architectural layers and threat persistence \\
    MCP-Guard~\cite{taxonomy250810991} & Attack types in tool interactions \\
    \bottomrule
  \end{tabularx}
\end{table}

\begin{table}[htbp]
  \caption{Academic sources for taxonomy construction, second group.}
  \label{tab:academic-sources-b}
  \centering
  \small
  \renewcommand{\arraystretch}{1.10}
  \begin{tabularx}{\columnwidth}{@{}>{\raggedright\arraybackslash}X@{\hspace{7pt}}>{\raggedright\arraybackslash}X@{}}
    \toprule
    Source & Classification perspective \\
    \midrule
    MCP-SafetyBench~\cite{taxonomy251215163} & Server, host, and user-side attacks \\
    MCPGuard~\cite{taxonomy251023673} & Hijacking, web, and supply-chain threats \\
    MCPXKIT~\cite{taxonomy250812538} & Injection and adversary interaction types \\
    MCP Security Bench~\cite{taxonomy251015994} & Attacks across the tool-use pipeline \\
    Agentic attack and defense survey~\cite{taxonomy260311088} & Agent attack surfaces and defenses \\
    Malicious Skills in the wild~\cite{taxonomy260206547} & Observed malicious Skill behaviors \\
    Skill vulnerabilities at scale~\cite{skillscan2026} & Skill vulnerability and risk categories \\
    Securing Agentic AI~\cite{taxonomy250419956} & Agent threat models and mitigations \\
    Agentic AI Security~\cite{taxonomy251023883} & Threats, defenses, and evaluation \\
    Emerging Skill ecosystem threats~\cite{taxonomy260528588} & Skill ecosystem attack scenarios \\
    Agent Skills for language models~\cite{taxonomy260212430} & Skill architecture and security concerns \\
    ClawHub Security Signals~\cite{clawhub2026} & Attack surfaces and scanner signals \\
    Agent Security Bench~\cite{taxonomy241002644} & Agent attack and defense categories \\
    When Safe Skills Collide~\cite{taxonomy260600448} & Cross-Skill compositional risks \\
    How Agents Ask for Permission~\cite{taxonomy260713718} & Permission specification and enforcement \\
    Agentic Skills beyond tool use~\cite{taxonomy260220867} & Skill lifecycle and security concerns \\
    Standardized AI threat taxonomy~\cite{taxonomy251121901} & Security, governance, and compliance \\
    \bottomrule
  \end{tabularx}
\end{table}

\begin{table}[htbp]
  \caption{Industry reports, standards, frameworks, and practitioner guidance.}
  \label{tab:industry-sources}
  \centering
  \small
  \renewcommand{\arraystretch}{1.10}
  \begin{tabularx}{\columnwidth}{@{}>{\raggedright\arraybackslash}X@{\hspace{7pt}}>{\raggedright\arraybackslash}X@{}}
    \toprule
    Source & Classification perspective \\
    \midrule
    ENISA threat landscape~\cite{enisa2020aithreat} & AI assets, threats, and lifecycle stages \\
    Stanford risk classifications~\cite{stanfordagenticrisk} & Data sensitivity, actions, and oversight \\
    Berkeley agentic risk profile~\cite{berkeley2026agenticprofile} & Agent-specific risks and governance controls \\
    CSA AI Controls Matrix~\cite{csa2026aicm} & AI security and governance controls \\
    EY agentic risk categories~\cite{ey2026agenticrisk} & Oversight, execution, and cascading risks \\
    Microsoft failure-mode taxonomy~\cite{microsoft2026agenticfailure} & Agent failure modes and mitigations \\
    360 agent security report~\cite{report360agentsafety} & Framework, ecosystem, and sandbox risks \\
    Skill supply-chain defense guide~\cite{uncletom2026skillsupplychain} & Attack points across the Skill lifecycle \\
    CSA agentic framework profile~\cite{csa2026agenticrmf} & Agent autonomy and risk-management controls \\
    MOANJU agent risk classification~\cite{moanju2026agentclassification} & Environment control and agent autonomy \\
    OWASP AI Exchange~\cite{owaspaiexchange} & AI threats, privacy, and controls \\
    MITRE ATLAS~\cite{mitreatlas} & Adversarial tactics and techniques \\
    NIST AI Risk Management Framework~\cite{nist2023airmf} & Trustworthiness and risk-management functions \\
    OWASP MCP Top 10~\cite{owaspmcptop10} & Protocol, tool, and context risks \\
    OWASP language-model Top 10~\cite{owaspllm} & Application vulnerabilities and consequences \\
    \bottomrule
  \end{tabularx}
\end{table}